\documentclass[oldversion]{aa}
\usepackage{txfonts}
\usepackage{graphicx}
\usepackage{natbib}
\bibpunct{(}{)}{;}{a}{}{,}
\begin{document}

\title{Observations and asteroseismological analysis of the rapid subdwarf B pulsator EC 09582-1137 \thanks{Based on observations collected at the European Organisation for Astronomical Research in the Southern Hemisphere, Chile (proposal ID 080.D-0192). This study made extensive use of the computing facilities offered by the Calcul en Midi-Pyr\'en\'ees (CALMIP) project and the Centre Informatique National de l'Enseignement Sup\'erieur (CINES), France.}
}
\author{
S.K. Randall \inst{1}
\and V. Van Grootel \inst{2}
\and G. Fontaine \inst{3}
\and S. Charpinet \inst{2}
\and P. Brassard \inst{3}
}

\institute{
ESO, Karl-Schwarzschild-Str. 2, 85748 Garching bei M\"unchen, Germany; \email{srandall@eso.org}
\and Laboratoire d'Astrophysique de Toulouse-Tarbes, Universit\'e de Toulouse, CNRS, 14 Avenue Edouard Belin, 31400 Toulouse, France
\and D\'epartement de Physique, Universit\'e de Montr\'eal, C.P. 6128, Succ. Centre-Ville, Montr\'eal, QC H3C 3J7, Canada
}
\date{Received date / Accepted date}

\abstract
{
We made photometric and spectroscopic observations of the rapidly pulsating subdwarf B star EC 09582-1137 with the aim of determining the target's fundamental structural parameters from asteroseismology. This analysis forms part of a long-term programme geared towards distinguishing between different proposed formation scenarios for hot B subdwarfs on the basis of their internal characteristics. So far, secure asteroseismic solutions have been computed for 9 of these pulsators, and first comparisons with results from evolutionary calculations look promising.

The new data comprise $\sim$ 30 hours of fast time-series photometry obtained with SUSI2 at the NTT on La Silla, Chile, as well as 1 hour of low-resolution spectroscopy gathered with EMMI, also mounted on the NTT. From the photometry we detected 5 independent harmonic oscillations in the 135$-$170 s period range with amplitudes up to 0.5\% of the mean brightness of the star. In addition, we extracted two periodicities interpreted as components of a rotationally split multiplet that indicate a rotation period of the order of 2$-$5 days. We also recovered the first harmonic of the dominant pulsation, albeit at an amplitude below the imposed 4 $\sigma$ detection threshold. The spectroscopic observations led to the following estimates of the atmospheric parameters of EC 09582-1137: $T_{\rm eff}$ = 34,806$\pm$233 K, $\log{g}$ = 5.80$\pm$0.04, and $\log{\rm N(He)/N(H)}$ = $-$1.68$\pm$0.06.  

Using the observed oscillations as input, we searched in model parameter space for unique solutions that present a good fit to the data. Under the assumption that the two dominant observed periodicities correspond to radial or dipole modes, we were able to isolate a well-constrained optimal model that agrees with the atmospheric parameters derived from spectroscopy. The observed oscillations are identified with low-order acoustic modes with degree indices $\ell$ = 0,1,2, and 4 and match the computed periods with a dispersion of 0.57 \%. Non-adiabatic calculations reveal all theoretical modes in the observed period range to be unstable, an important a posteriori consistency check for the validity of the optimal model. The inferred structural parameters of EC 09582-1137 are $T_{\rm eff}$ = 34,806 K (from spectroscopy), $\log{g}$ = 5.788$\pm$0.004, $M_{\ast}$ = 0.485$\pm$0.011 $M_{\odot}$, $\log{(M_{env}/M_{\ast})}$ = $-$4.39$\pm$0.10, $R$ = 0.147$\pm$0.002 $R_{\odot}$, and $L$ = 28.6$\pm$1.7 $L_{\odot}$. We additionally derive the absolute magnitude $M_V$ = 4.44$\pm$0.05 and the distance $d$ = 1460$\pm$66 pc. 

}

\keywords{}
\titlerunning{Asteroseismology of EC 09582-1137}
\authorrunning{S.K. Randall et al.}
\maketitle

\section{Introduction}    

Understanding the formation of subdwarf B (sdB) stars is one of the remaining challenges related to stellar evolution theory. It is generally accepted that the progenitors of these hot, compact objects (20,000 $\lesssim T_{\rm eff} \lesssim$ 40,000 K and 5.0 $\lesssim \log{g} \lesssim$ 6.2) lose too large a fraction of their envelope mass near the tip of the first red giant branch (RGB) to ignite hydrogen-shell burning during the asymptotic giant branch (AGB) phase. They instead settle on the extreme horizontal branch (EHB), where they spend around 10$^8$ years as helium-core burning objects surrounded by an inert hydrogen-rich envelope before joining the white dwarf cooling track. However, the circumstances leading to the very fine-tuned mass loss necessary to produce an sdB star are still under debate. Formation channels proposed feature single star and binary evolution in different flavours, including increased mass loss due to helium-enrichment (e.g. \citealt{sweigart1997}), the ``hot flasher'' scenario (e.g. \citealt{d'cruz1996}), common envelope ejection, stable Roche lobe overflow, and the merger of two helium white dwarfs (modelled in detail by \citealt{han2002,han2003}). In principle, the accuracy and relative importance of these evolutionary channels can be tested by comparing the distributions of binary and structural parameters derived from the different simulations to those observed for sdB stars. While the binary properties can be measured using well established techniques like radial velocity measurements from spectroscopy, certain structural parameters such as the mass and internal compositional layering  can normally not be inferred using traditional methods. The former may be derived in the case of an eclipsing binary, although such systems are statistically rare. They can, however, be used to independently verify the mass estimates from asteroseismic analysis, as done very successfully for the case of PG 1336-018 \citep{maja2007,charp2008}. For its part, the determination of the internal layering, in particular the thickness of the hydrogen-rich envelope, is a pure product of asteroseismology. 

\begin{figure*}[t]
\centering
\includegraphics[width=10cm,angle=270,bb=80 130 550 720]{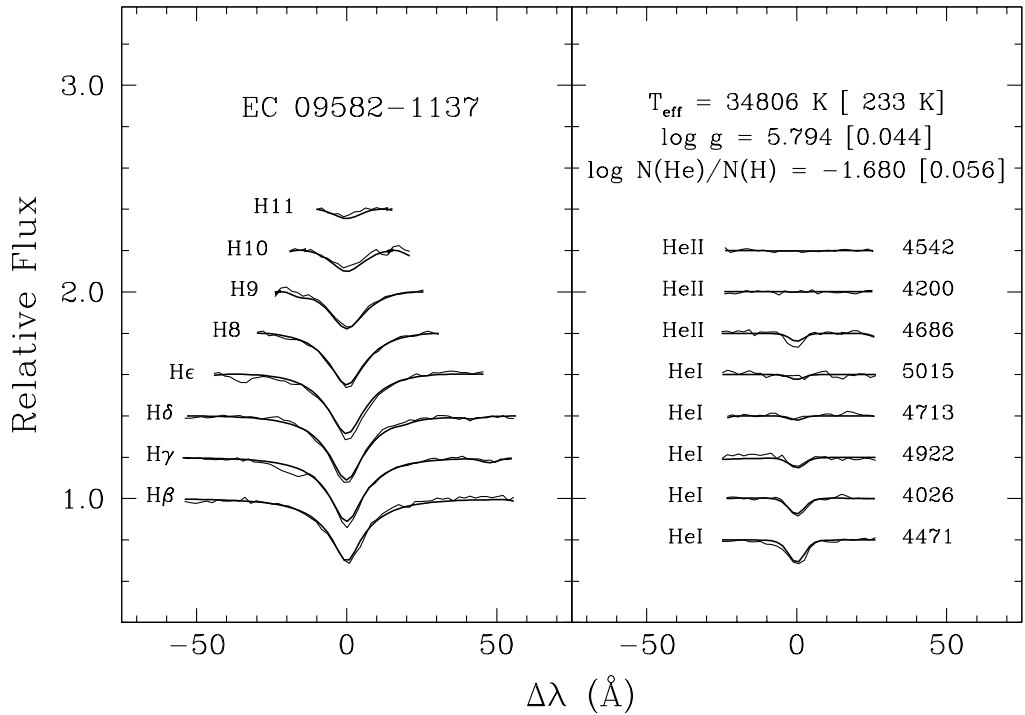}
\caption{Available H and He lines from the combined EMMI spectrum of EC 09582-1137, overplotted with the best model atmosphere fit. The atmospheric parameters derived are indicated in the plot.}
\label{spec}
\end{figure*}

The discovery of rapid nonradial pulsators among subdwarf B stars \citep{kilkenny1997} opened up the exciting possibility of using asteroseismology to probe the interiors of these objects and determine the internal parameters. Often referred to as EC 14026 stars after the prototype, the rapid sdB pulsators exhibit luminosity variations on typical timescales of 100$-$200 s with amplitudes of a few milli-magnitudes (mmags) and are found near the hot end of the EHB at 29,000 $\lesssim T_{\rm eff} \lesssim$ 36,000 K. The oscillations have been modelled as $p$(pressure)-mode instabilities driven by a classical $\kappa$-mechanism associated with an iron opacity peak arising from the ionisation of K-shell electrons \citep{charp1996}. It is noteworthy that models assuming a uniform iron abundance in the envelope cannot excite pulsations unless the latter is artificially boosted to several times the solar value, an unrealistic proposition for these metal-poor stars. The problem was overcome by including radiative levitation and the resulting non-uniform iron abundance profiles in the so-called ``second-generation'' models \citep{charp1997}. These are able to explain the pulsations observed in real EC 14026 stars to the point where the location of the instability strip can be accurately reproduced and the quantitative asteroseismological interpretation of period spectra for individual stars has become possible (see \citealt{fontaine2008,ostensen2009} for recent reviews on the subject). To date, full asteroseismic analyses have been carried out for 9 rapidly pulsating sdB  stars: PG 0014+067 \citep{brassard2001,charp2005c,brassard2008}, PG 1047+003 \citep{charp2003}, PG 1219+534 \citep{charp2005a}, Feige 48 \citep{charp2005b,val2008a}, EC 20117-4014 \citep{randall2006c}, PG 1325+101 \citep{charp2006}, PG 0911+456 \citep{randall2007}, Balloon 090100001 \citep{val2008b} and PG 1336-018 \citep{charp2008}. In this Paper we present the tenth asteroseismological analysis of an EC 14026 star.

The target of the present study, EC 09582-1137, was identified in the Edinburgh-Cape (EC) survey zone I paper \citep{kilkenny1997b} as a subdwarf B star with no spectroscopic signature of a cooler companion. Infrared measurements from the 2MASS database (see the web page www.ipac.caltech.edu/2mass) tend to confirm this, although their accuracy is rather poor because of the relative faintness of the star ($V$=15.26 from the EC survey). 
EC 09582$-$1137 was discovered to be pulsating only relatively recently by \citet[][hereafter K2006]{kilkenny2006}, who uncovered two periodicities at 136.0 and 151.2 s with amplitudes of 8.3 and 7.5 mmags respectively. No further oscillations were detected down to a threshold of 1 mmag.     

In what follows, we present new spectroscopic and photometric observations of EC 09582-1137. These are used  to derive atmospheric parameters and to detect oscillation frequencies beyond those already known (Section 2). We then provide details of the asteroseismological analysis conducted on the basis of our observations, and discuss the internal parameters inferred (Section 3). Finally, we summarise our results and put them into the context of current research in Section 4.

\section{Observations \& analysis}

\subsection{Spectroscopy and atmospheric analysis}

We were allocated one hour of service mode observing time with EMMI at the 3.5-m NTT located at La Silla, Chile (for details on the instrument see the web page http://www.eso.org/sci/facilities/lasilla/instruments/emmi). Three low-resolution spectra, each with an integration time of 1000 s, were obtained on 24 December 2007 using Grating 4. Each spectrum was bias and flat field corrected, extracted, sky subtracted and wavelength calibrated using standard IRAF routines. The combined spectrum has a wavelength resolution $\rm \Delta\rm \lambda\sim$ 5 $\rm \AA$ and signal-to-noise of S/N $\sim$ 160/pixel at the central wavelength $\rm \lambda_C$= 4355 \AA, and covers the wavelength range from 3413 to 5278 \AA. It was analysed using a detailed grid of non-LTE model atmospheres and synthetic spectra designed especially for subdwarf B stars. These were computed using the public codes TLUSTY and SYNSPEC \citep{hubeny1995,lanz1995}, and include helium but no metals. Fig. \ref{spec} shows the best simultaneous model fit to the available Balmer and helium lines, and indicates the atmospheric parameters inferred. The dip near $H\gamma$ is not understood and interpreted as a glitch in the data, of unknown origin. As typical in sdB stars with such parameters, the HeII 4686 line is observed stronger than predicted, which is likely due to the lack of metals in our models \citep[see][]{heber2000}. Note that for sdB stars around $T_{\rm eff}\sim$ 35,000 K this does not severely affect the atmospheric parameters derived \citep{geier2007}. Comparing the latter with e.g. Fig. 1 of \citet{fontaine2008} reveals EC 09582-1137 to be a typical EC 14026 star in terms of atmospheric parameters, located towards the hot end of the instability strip.

\subsection{Time-series photometry and frequency analysis}

For the photometry part of the observations we were allocated 5 nights with SUSI2 (SUperb Seeing Imager), mounted on the NTT at La Silla. The observing log is given in Table \ref{obslog}. The full dataset comprises $\sim$ 30 hours of time-series photometry spread over 103 hours, which yields a frequency resolution of 2.7 $\mu$Hz. Given the relative faintness of the star, we chose to keep the filter slot empty with the aim of collecting as many photons as possible. In order to reduce the readout time we employed the SUSI2 'windowing' mode, which reads out only a pre-defined part of the chip, and moreover used 3$\times$3 binning. Thanks to the small pixel scale of the SUSI2 imager (0.08\arcsec/pixel without binning), the FWHM of the stars in the images was always above $\sim$ 3 pixels, even during periods of good seeing (the seeing on the images ranges from 0.7\arcsec to 4\arcsec, with an average of 1.3\arcsec). The resulting dead time between exposures is 13 s, which together with the exposure time of 10 s gives a cycle time of 23 s.

\begin{table}[h]
\caption{Photometry obtained for EC 09582-1137 (2008)}
\label{obslog}
\centering
\begin{tabular}{c c c}
\hline\hline
Date (UT) & Start time (UT) & Length (h) \\
\hline 
10 Mar & 00:52 & 00:51 \\
10 Mar & 06:27 & 01:23 \\
11 Mar & 03:55 & 04:20 \\
12 Mar & 00:40 & 07:11 \\
13 Mar & 00:17 & 07:44 \\
14 Mar & 00:19 & 07:41 \\
\hline
\end{tabular}
\end{table}

\begin{figure}[t]
\centering
\includegraphics[width=7.0cm,angle=270,bb=20 60 560 700]{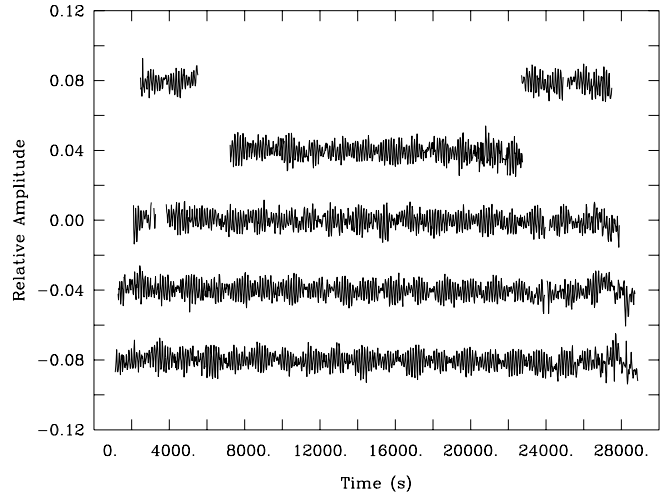}
\caption{All light curves obtained for EC 09582-1137 with SUSI2. The data have been shifted arbitrarily along the $x$ and $y$ axes for visualisation purposes. From top to bottom the curves refer to the nights of 10, 11, 12, 13 and 14 March 2008. Details are given in Table \ref{obslog}.}
\label{curves}
\end{figure}

\begin{figure}[t]
\centering
\includegraphics[width=7.0cm,angle=270,bb=20 60 560 700]{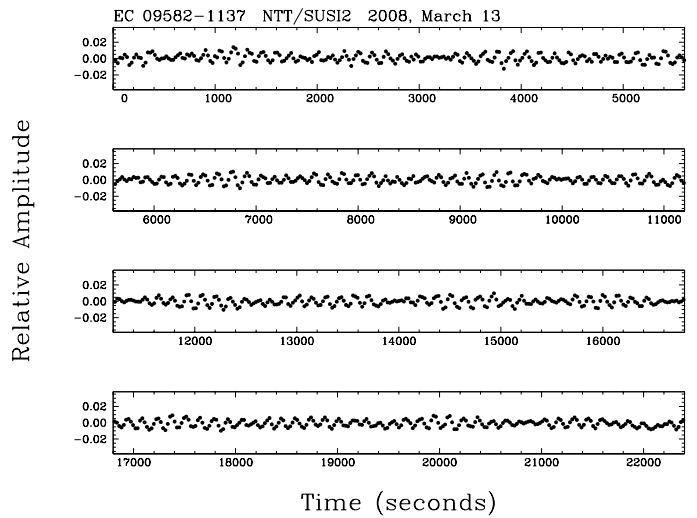}
\caption{Expanded view of a portion of the light curve obtained on 13 March 2008 in units of fractional brightness intensity and seconds.}
\label{12mar}
\end{figure}

The data were reduced using standard IRAF aperture photometry routines, except that the photometric aperture size was adjusted to 2.75 times the FWHM in each image. Differential photometry was obtained on the basis of 3 suitable comparison stars of similar brightness to the target that were specifically included in the windowed area of the CCD chip. The resulting light curves for all 5 nights are displayed in Figure \ref{curves}, where the large gaps in the top two light curves are due to the telescope being closed because of high humidity. It is apparent that the longest two light curves corresponding to the last two nights are degraded in quality towards the end; this is a result of the high airmass ($X\sim$ 2.3) at which the setting target was observed. An expanded version of one of the light curves is shown in Figure \ref{12mar}, where the pulsation and beating between individual frequencies can be better appreciated.

\begin{figure}[t]
\centering
\includegraphics[width=7.0cm,bb=80 90 510 780]{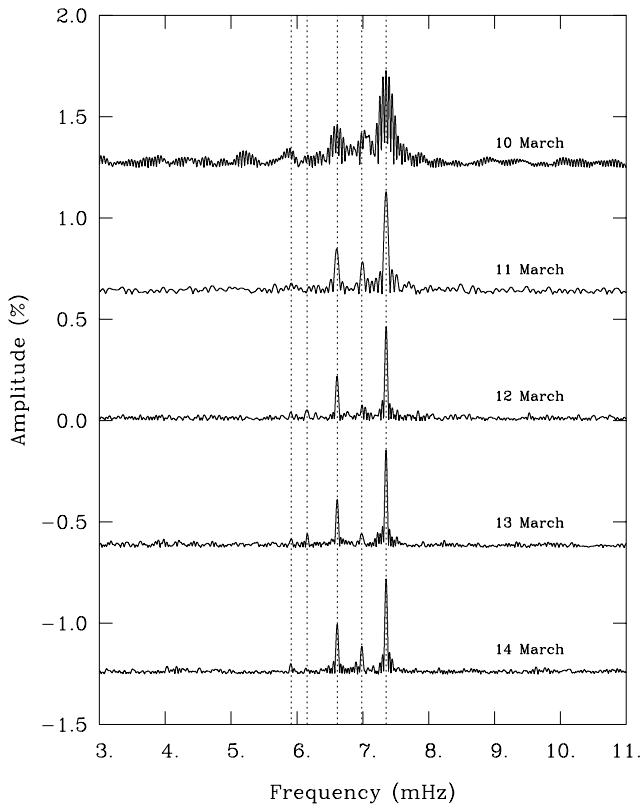}
\caption{Fourier transforms for the individual light curves for each night in the 3$-$11 mHz range. The curves have been shifted arbitrarily along the $y$-axis for visualisation purposes, however the amplitude scale is the same as for the original Fourier transform. The locations in frequency space of the 5 independent harmonic oscillations extracted down to a threshold of 4 $\sigma$ are marked by the dotted vertical lines.
}
\label{fts}
\end{figure}

\begin{figure}[b]
\centering
\includegraphics[width=8.0cm,bb=100 180 500 680]{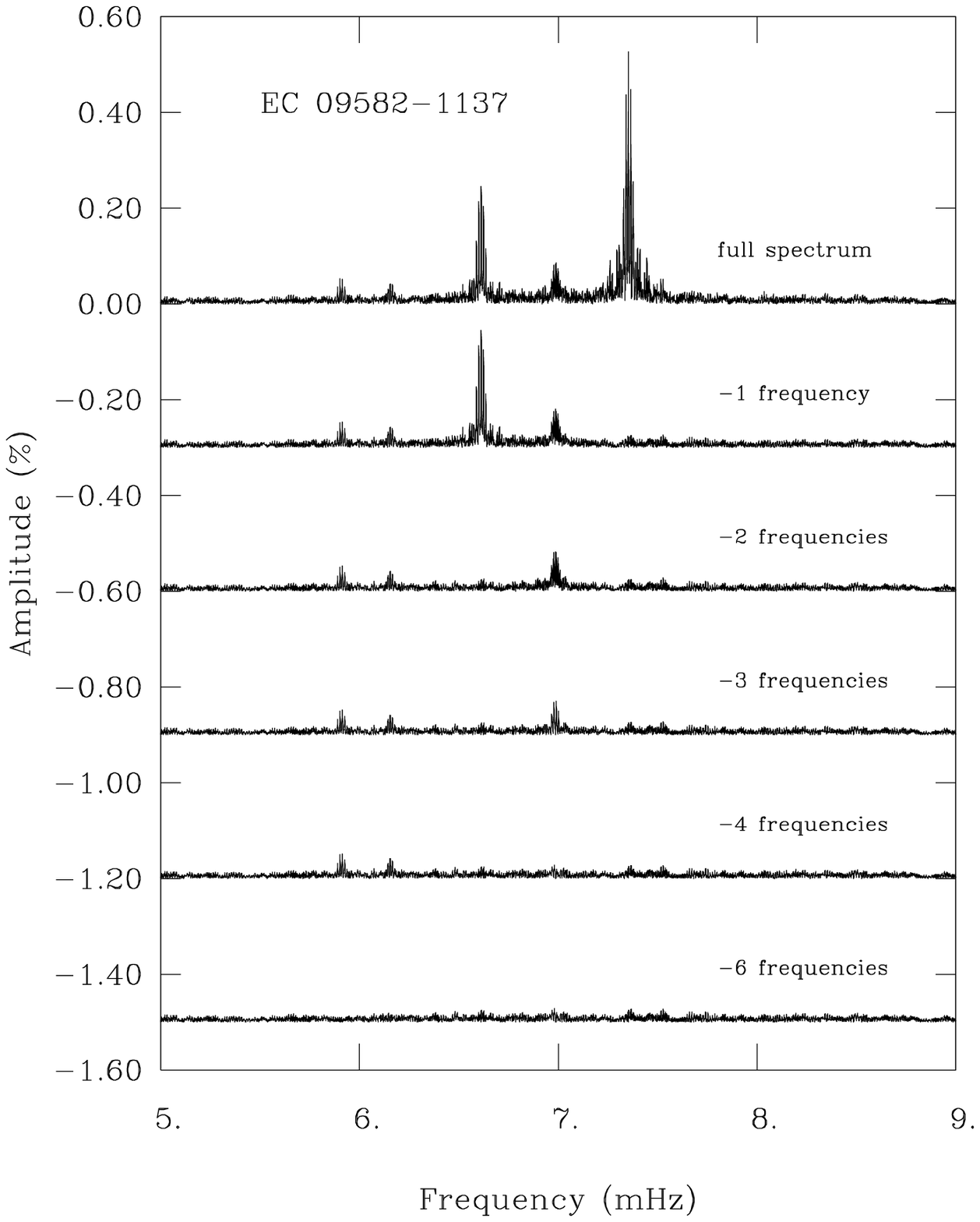}
\caption{Fourier transform of the entire data set zoomed in to the 5$-$9 mHz range (the spectrum outside this range is consistent with noise). The lower transforms show the successive steps of prewhitening as indicated for frequencies detected above 4 $\sigma$. They have been shifted arbitrarily along the $y$-axis for visualisation purposes, however the amplitude scale is the same as for the original Fourier transform.
}
\label{prewhitening}
\end{figure}

We computed the Fourier transforms (FT) for each of the light curves individually as well as for different combinations of data sets. The individual nightly Fourier transforms are plotted in Figure \ref{fts}. In the end, the lowest noise level was achieved by combining the entire data set, which is not all that surprising considering the relative homogeneity of the data quality. The resulting Fourier transform is displayed in the top panel of Figure \ref{prewhitening} in the zoomed-in 5$-$9 mHz range, the remaining power spectrum out to the Nyquist frequency of 21.7 mHz being consistent with noise. The lower curves refer to successively pre-whitened data, as indicated. During the process of pre-whitening, the dominant peak is identified from the FT and used as input for a least-squares (LS) fit to the light curve in which the amplitude and phase are determined. The resulting sinusoid is then subtracted from the light curve, and the sequence is repeated until all periodicities down to a specified amplitude threshold have been extracted. The subtraction of periodicities in time rather than frequency space ensures that not only the actual FT peak, but also the sidelobes caused by the daily gaps in the data are removed in each successive FT.  

 We detected a total of 6 frequencies down to an imposed threshold of 4 times the average noise level. In Table \ref{freqs} we list these frequencies together with the corresponding periods, amplitudes and phases, as well as the S/N derived from the ratio of the amplitude to the 1$\sigma$ noise level (measured to average 0.0071 \% over the 0$-$16 mHz frequency range). The periodicities are numbered according to amplitude rank, and the corresponding frequencies and amplitudes reported by K2006 are indicated where applicable. We also include two lower amplitude periodicities (marked by asterisks in the S/N column): the probable multiplet component $f_3+$ (see below), and the harmonic of the dominant frequency $f_1H$. While the harmonic has a S/N of only 2.3, and as such falls below the threshold for credible peaks by any standards, it is found at {\it exactly} twice the frequency of the dominant oscillation, and is thus likely to be real. It has no part to play in the asteroseismological analysis detailed below, however it is interesting to note that we detect it for this sdB star despite the lack of significant convection in the envelope, which is sometimes invoked to explain the presence of harmonics in the pulsation spectra of other types of stars \citep{wu2001}. 

The errors on the period (frequency), amplitude, and phase were calculated using the recipes described in \citet[][hereafter MO]{montgomery1999}. These give uncertainties on the amplitudes and phases virtually identical to those obtained from the LS fit to the light curve, at least when applied to the full data set. When considering one night at a time, the more conservative MO recipes yield uncertainties around 10 \% larger, and it is these estimates that we employed for the modelling of the amplitude and phase variations discussed below. In addition to the 8 frequencies listed in Table \ref{freqs}, two further low amplitude periodicities were found just above the less stringent threshold of 3$\sigma$ at 135.750 s ($A \sim$ 0.022 \%) and 132.826 s ($A \sim$ 0.022 \%). These are not considered reliable enough as input for the initial asteroseismic analysis, but are nevertheless interesting to note.     

\begin{table*}[t]
\caption{Oscillations detected from the combined light curve for EC 09582-113.}
\label{freqs}
\centering
\begin{tabular}{l c c c c c c c}
\hline\hline
Rank & Period & Frequency & Amplitude & Phase & S/N & $f_{K2006}$ & $A_{K2006}$\\
{} &  (s) & (mHz) &  (\%) & (s) & {} & (mHz) &  (\%)\\
\hline 
$f_4$ &  169.1109$\pm$0.0054 & 5.91328$\pm$0.00019 & 0.0445$\pm$0.0056 &
49.02$\pm$3.38  & 6.3 & ... & ... \\
$f_5$ &  162.4811$\pm$0.0060 & 6.15456$\pm$0.00023 & 0.0371$\pm$0.0056 &
107.41$\pm$3.90 & 5.2 & ... & ...\\
$f_2$ & 151.2420$\pm$0.0008 & 6.61192$\pm$0.00004 & 0.2297$\pm$0.0056 &
100.26$\pm$0.59 & 32.4 & 6.61254 & 0.69 \\
$f_3-$ &  143.2468$\pm$0.0067 & 6.98096$\pm$0.00033 & 0.0258$\pm$0.0056 &
10.46$\pm$4.95  & 3.6$^\ast$ & ... & ...\\
$f_3$ & 143.1418$\pm$0.0024 & 6.98608$\pm$0.00012 & 0.0726$\pm$0.0056 &
14.99$\pm$1.75  & 10.2 & ... & ...\\
$f_3+$ & 143.0664$\pm$0.0026 & 6.98976$\pm$0.00013 & 0.0663$\pm$0.0056 &
63.31$\pm$1.92  & 9.3 & ... & ...\\
$f_1$ & 135.9967$\pm$0.0003 & 7.35312$\pm$0.00002 & 0.4739$\pm$0.0056 &
14.67$\pm$0.26  & 66.7 & 7.35342 & 0.76\\
$f_1H$ &  67.9970$\pm$0.0024 & 14.70653$\pm$0.00052 & 0.0163$\pm$0.0056 &
19.29$\pm$3.71 & 2.3$^\ast$ & ... & ...\\
\hline
{\footnotesize $^{\ast}$ below 4$\sigma$}
\end{tabular}
\end{table*}
 
Examining the periodicities listed in Table \ref{freqs}, we find that the period spectrum of EC 09582-1137 is dominated by two peaks at $\sim$136 s and $\sim$151 s respectively. These have amplitudes an order of magnitude higher than the other oscillations, and correspond to the two frequencies detected by K2006. However, the amplitudes of both, and in particular the secondary peak are much higher in the earlier data. This could be partly due to the slightly different QE curves of the SUSI2 CCD and the UCT-CCD used by K2006, however upon comparison these were found to be quite similar. In any case, the dramatic relative change in power between the two peaks cannot be explained in terms of different CCD responsitivities. It is more likely that the amplitudes have changed either intrinsically or as a result of the beating between unresolved modes (our data have a longer baseline). The weaker oscillations fall below the K2006 detection threshold of 0.09 \%, so it is impossible to say whether their amplitudes have changed or not. Looking at the frequencies of the two main peaks, these seem to be systematically lower by $\sim$ 0.5 $\mu$Hz in our data. It is unclear whether this is a coincidence as the potential frequency shift is similar to the accuracy of the measurements, or whether it could be a tentative manifestation of radial velocity variations caused by the presence of an unseen companion.

\begin{figure}[b]
\centering
\includegraphics[width=7.0cm,angle=270,bb=70 70 560 700]{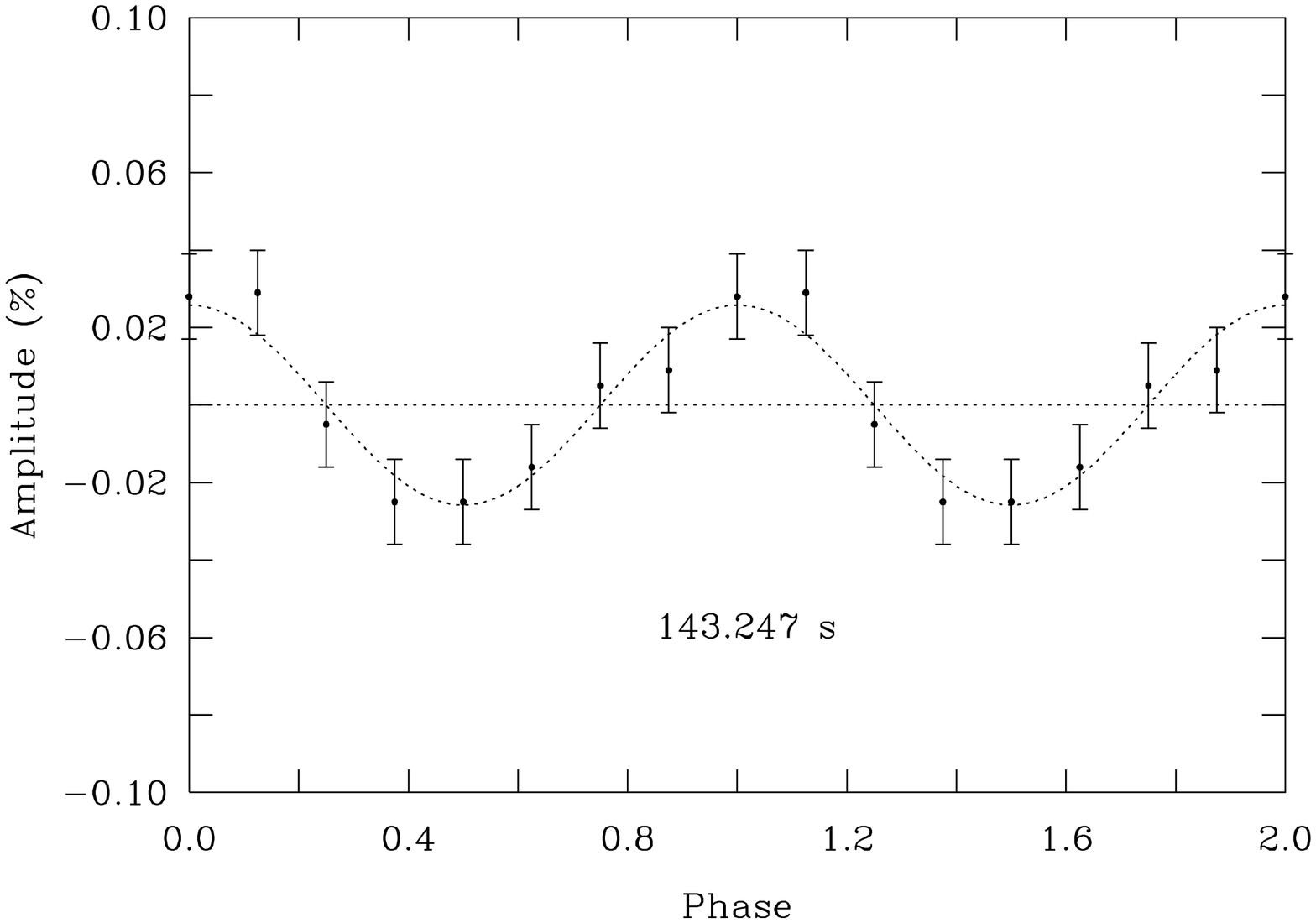}
\caption{The combined SUSI2 light curve folded on the 143.247 s periodicity and prewhitened of the other 7 frequencies listed in Table \ref{freqs}.
}
\label{folded}
\end{figure}

\begin{figure}[h]
\centering
\includegraphics[width=7.0cm,bb=100 100 530 680]{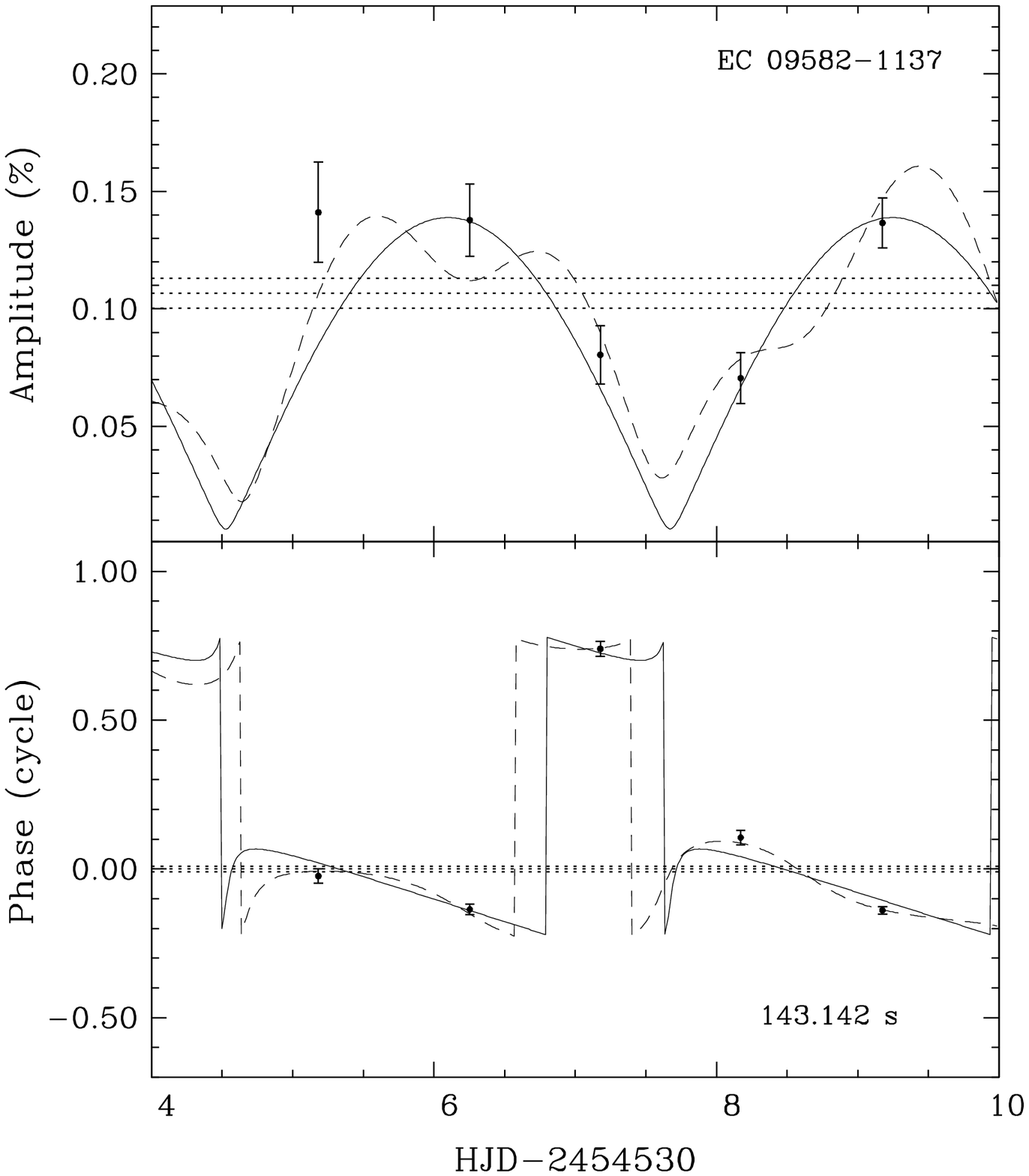}
\caption{Nightly variation in amplitude (top panel) and phase (bottom panel) of the 143 s complex (dots with error bars). The dashed (continuous) curves show the amplitude and phase variations as simulated for the beating action of the triplet structure $f_3-,f_3,f_3+$ (doublet structure $f_3,f_3+$). The amplitude/phase and the associated error derived for the full data set is indicated by the horizontal dotted line.  
}
\label{ampvar}
\end{figure}  

An interesting feature of the period spectrum detailed in Table \ref{freqs} is the apparent multiplet structure around $f_3$. The two higher frequency components $f_3$ and $f_3+$ have comparable amplitudes well above the 4$\sigma$ limit, while $f_3-$ falls below the threshold. However, we nevertheless believe that it corresponds to a sinusoidal variation in the data rather than a spurious peak. In order to convince the reader of this, Fig. \ref{folded} shows the light curve prewhitened by the 7 other oscillations of Table \ref{freqs} and folded onto the 143.247 s component. The individual data points have been combined into 8 phase bins for clarity, and the sinusoid derived from the least-squares fit to the light curve has been overplotted. It is clear that, within the estimated errors, the points provide a good fit to the sinusoid, as they should if the peak is real.

The most commonly invoked explanation for closely spaced frequency multiplets in subdwarf B stars is rotational splitting. It is well known that in a rotating star spherical symmetry is broken and the $m$-fold degeneracy of a mode characterised by degree $\ell$ and radial order $k$ is lifted. Treating rotation as a first-order perturbation under the assumption of solid-body rotation results in a frequency splitting between adjacent values of the azimuthal index $m$ (where $m=-\ell,(-\ell+1),...,(\ell-1),\ell$) of
\begin{equation}
\Delta f \simeq \frac{1-C_{kl}}{P_{rot}}
\end{equation}
where $P_{rot}$ is the rotation period and $C_{kl}$ is the dimensionless first-order rotation coefficient. The latter is normally negligible compared to unity for the $p$-modes observed in EC 14026 stars (see Table \ref{perfit} for the example of the optimal model).

In the case of EC 09582-1137, the $f_3$ multiplet is not evenly spaced in frequency, with $\Delta f_3,f_3-=5.12\pm 0.35$ $\mu$Hz and $\Delta f_3,f_3+=3.68\pm 0.18$ $\mu$Hz. We believe that this is related to the multiplet not being completely resolved, even in the full data set. Indeed, the frequency resolution of the combined photometry is 2.7 $\mu$Hz, of the same order of magnitude as the frequency splitting measured, and when analysing data subsets of 2$-$4 continuous nights we found $\Delta f$ to decrease as the time baseline increased. Going back to the Fourier transforms for the individual nights (see Fig. \ref{fts}), we find that the 143 s structure manifests itself as one broad peak that appears to significantly change its amplitude from one night to the next. To investigate this in more detail, we show the nightly variation in amplitude and phase of the (unresolved) 143 s peak in Fig. \ref{ampvar}. The overplotted dashed (continuous) curves refer to the amplitude and phase variations expected from the beating action of the triplet structure $f_3-,f_3,f_3+$ (doublet structure $f_3,f_3+$). While both the doublet and triplet scenario can account for the amplitude variations observed relatively well, the triplet model is slightly more successful at matching the observed phase variations. The agreement in amplitude is by no means perfect, particularly for the first two nights of observation, where the data sets were somewhat shorter than for the last three nights. We interpret this in terms of an insufficient resolution of the dataset: when extracting the nightly amplitudes and phases of the 143 s structure, we kept the period fixed at the $f_3$ value listed in Table \ref{freqs}. If that period is poorly measured because of insufficient resolution, this will induce small errors in the nightly amplitude and phase measurements. Nevertheless, we believe that, together with Fig. \ref{folded}, the results of Fig. \ref{ampvar} point towards the lower amplitude $f_3-$ component being real, and the 143 s complex containing at least a triplet of peaks. In this case one could infer an approximate rotational period of $P_{rot}\sim$2.63 days based on the mean frequency spacing of $\Delta f$=4.4 $\mu$Hz. However, given the limited resolution of our dataset it is of course quite possible that the structure is in fact an unresolved quintuplet. In this case we may be seeing the $m$=-2, $m$=0, and $m$=+2 components of an $\ell$=2 mode, the frequency spacing would correspond to twice the splitting from rotation, and the rotation period of the star would be 5.26 days.   

 It is unfortunate that we do not detect rotational splitting in any of the other peaks, as this would allow us to determine the rotational period with more certainty. While it is quite likely that one of the dominant periodicities is a radial mode and as such would not be subject to rotational splitting, at least some of the other modes must have higher degree indices. For the lower amplitude modes $f_4$ and $f_5$ the multiplet components probably fall far below our detection threshold, but for $f_1$ and $f_2$ they could be measurable, assuming of course that these are {\it not} radial modes. In this context, the 3$\sigma$ oscillation at 135.750 s becomes interesting as a potential multiplet to the dominant 135.997 s periodicity. However, the frequency splitting is $\Delta f$=13.5 $\mu$Hz, much larger than that attributed to rotational splitting from the $f_3$ multiplet, so we believe it to correspond to an independent harmonic oscillation with no direct connection to the strongest peak. Considering the case of $f_2$=151.242 s, we indeed find a very closely spaced component just below the 3$\sigma$ level at 151.206 s, which would imply a frequency separation of $\Delta f$=1.5 $\mu$Hz. Since the latter is below the formal resolution of the dataset it cannot be taken at face value, but it does hint at the presence of some multiplet structure, quite possibly with a spacing around half that found for $f_3$. This would then imply $\ell\geq$ 2 for $f_3$, however a longer data set is needed for confirmation.  

In summary, we detected 5 independent harmonic oscillations as well as one peak presumed to constitute a rotationally split multiplet component above the 4$\sigma$ threshold in our data. Imposing a lower threshold of 3 $\sigma$ we additionally find a second split multiplet component and two lower amplitude frequencies. While the multiplet component forms part of a triplet and is thought to be real, the weaker independent oscillations are thought of as insecure detections. Finally, we uncovered the first harmonic of the dominant oscillation at a very low amplitude. 

%\begin{figure}
%\centering
%\includegraphics[width=8.5cm,bb=100 200 500 650]{checkft.ps}
%\caption{Comparison between the Fourier transform of the entire data set (Full spectrum) and that reconstructed on the basis of the 6 periods, amplitudes and phases derived from least squares fitting to the light curves (Model spectrum). The lower curve shows the point-by-point difference between the actual FT and the model FT.
%}
%\label{checkft}
%\end{figure}

\section{Asteroseismic analysis}

\subsection{Methodology}

For the asteroseismological analysis of EC 09582-1137 we follow the well-known forward method described in detail by \citet{charp2005a}. This approach is based on the requirement of {\it global} optimisation, implying that {\it all} the periods observed are {\it simultaneously} matched to those computed from sdB star models. For the latter we employ our so-called ``second-generation'' (2G) models (see \citealt{charp1996,charp2001}), static structures composed of a hard ball nucleus surrounded by a more accurately modelled envelope extending down to a logarithmic depth of $\log{q}\equiv\log{(1-M(r)/M_{\ast})}\simeq -0.05$, sufficient for an accurate computation of the shallow $p$-modes observed in the EC 14026 stars. An important feature of the 2G models is that they incorporate microscopic diffusion under the assumption of an equilibrium having been established between gravitational settling and radiative levitation. For the case of the sdB star models it is iron that is assumed to be levitating in a pure hydrogen background. Since it is the local overabundance of iron in the driving region that creates the opacity bump necessary for the excitation of pulsations through the $\kappa$-mechanism, including diffusion processes is vital when it comes to predicting whether a mode is unstable or not. Moreover, microscopic diffusion changes the stellar structure sufficiently to affect the frequencies of pulsations and must therefore be taken into account when attempting quantitative analyses \citep{fontaine2006b}. The 2G models are specified by four fundamental input parameters: the effective temperature $T_{\rm eff}$, the surface gravity $\log{g}$, the total stellar mass $M_{\ast}$, and the fractional thickness of the hydrogen-rich envelope $\log{q(\rm H)}=\log{M(\rm H)/M_{\ast}}$. The latter parameter is intimately related to the more commonly used quantity $M_{env}$, which corresponds to the total mass of the H-rich envelope\footnote{Note that the parameter $M_{env}$ commonly used in extreme horizontal branch stellar evolution includes the mass of hydrogen contained in the thin He/H transition zone, whereas the parameter $M(H)$ used in our envelope models does not. They can be related with $\log{[M_{env}/M_{\ast}]}=\log{[M(H)/M_{\ast}]}+C$, where $C$ is a small positive term slightly dependent on the model parameters that can be computed from the converged model using the mass of hydrogen present in the transition zone itself.}. 

The 2G models are the input for the computation of adiabatic and non-adiabatic oscillation modes using two efficient codes based on finite element techniques \citep{brassard1992,fontaine1994}. The first solves the four adiabatic oscillation equations and constitutes an intermediate but necessary step to derive estimates for the periods and eigenfunctions that are then used as first guesses for the solution of the full, non-adiabatic set of oscillation equations. The latter computes a number of quantities for each mode, including some that can be directly compared with observed quantities, such as the period $P_{th}(=2\pi/\sigma_R$, where $\sigma_R$ is the real part of the complex eigenfrequency), and the stability coefficient $\sigma_I$ (the imaginary part of the eigenfrequency). If $\sigma_I$ is positive, the mode is stable, while if it is negative the mode is excited, and may therefore be observable if its amplitude is large enough. Although the asteroseismological analysis is based on the more accurately computed adiabatic modes, the non-adiabatic approach can be used as a consistency check in the sense that the periods observed should also be predicted to be unstable. As is standard, the pulsations are computed assuming perfect spherical symmetry, which is well justified for slowly rotating stars such as EC 09582-1137. This implies that each theoretical mode is (2$\ell$+1)-fold degenerate in eigenfrequency, and all multiplet components seen in the observed spectrum (such as the three $f_3$ components identified in the present case) must be considered as a single harmonic oscillation for the purpose of the asteroseismological analysis. In this context, a pulsation mode is then completely defined by the degree index $\ell$ and the radial order $k$. 

\begin{figure*}[t]
\begin{tabular}{cc}
{\includegraphics[width=8.8cm]{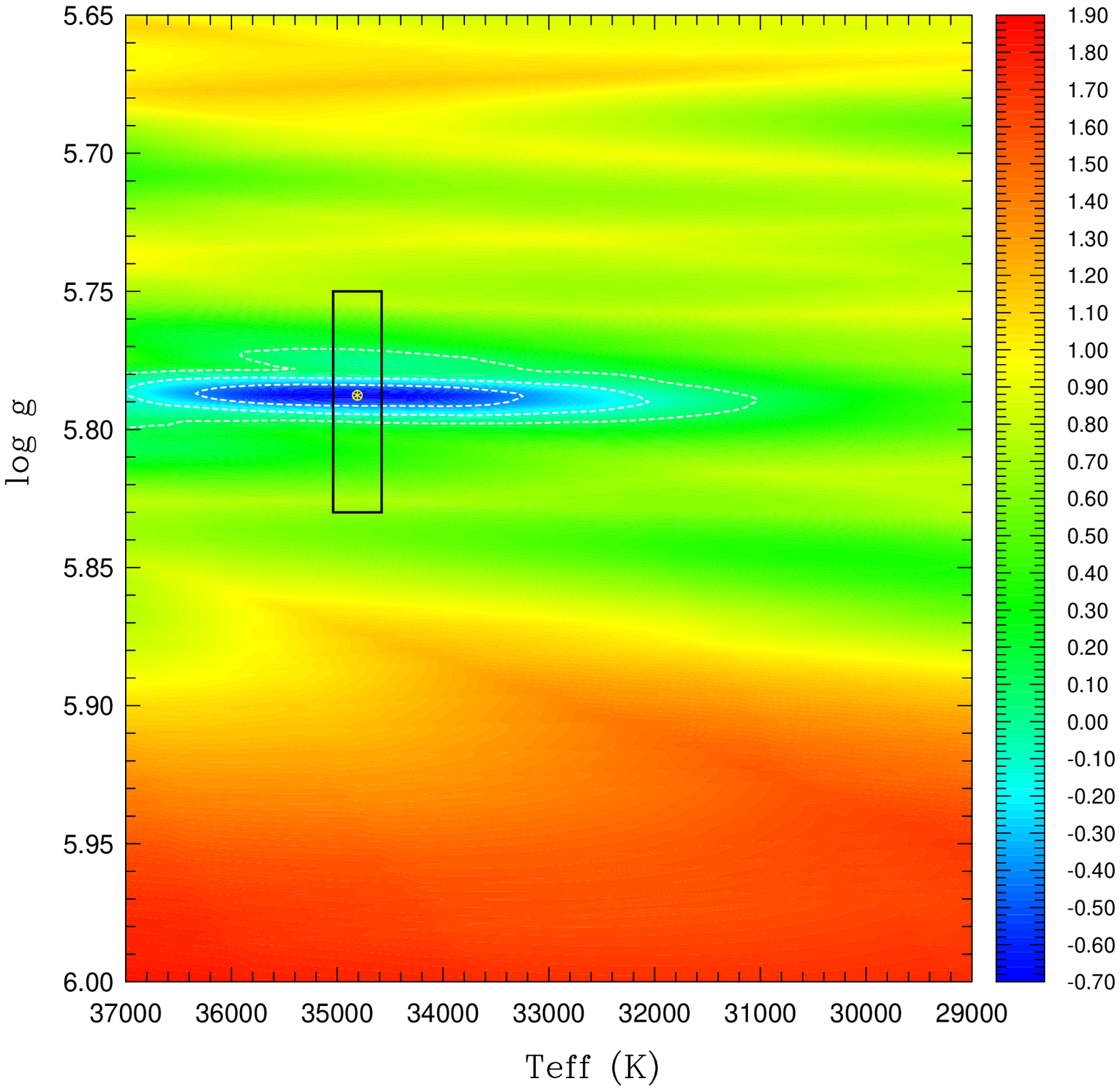}} & {\includegraphics[width=8.8cm]{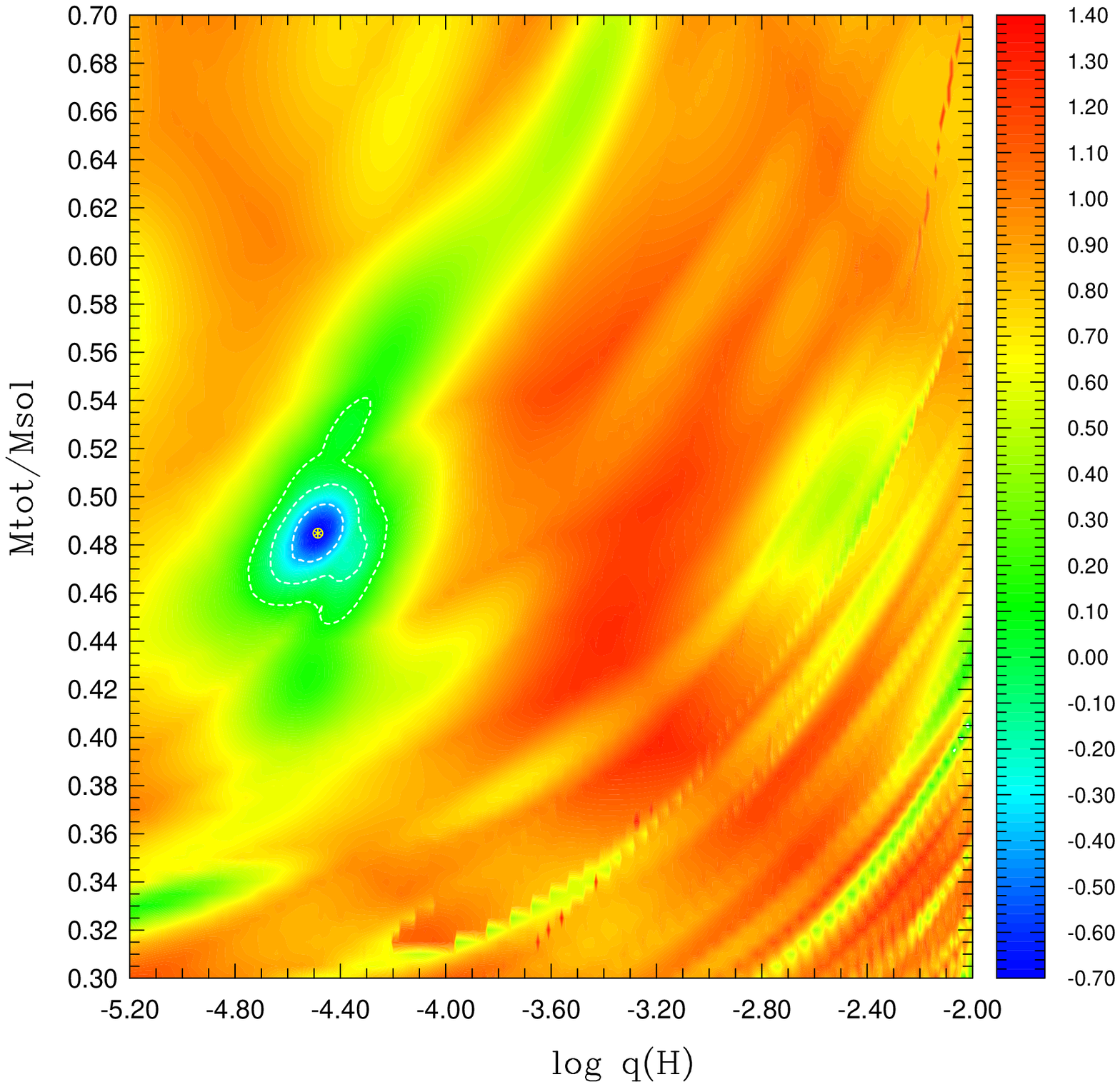}} \\
\end{tabular}
\caption{{\it Left panel:} Slice of the $S^2$ function (in logarithmic units) along the $T_{\rm eff}-\log{g}$ plane with $M_{\ast}$ and $\log{q(\rm H)}$ set to their optimal values of $M_{\ast}$=0.49 and $\log{q(\rm H)}$=$-$4.48. The solid rectangle represents our spectroscopic estimate of the atmospheric parameters for EC 09582-1137. Also indicated are the 1,2 and 3$\sigma$ limits on the $S^2$ minimum (white dashed contours). The errors on the asteroseismologically derived parameters are estimated from the semi-axes of the 1$\sigma$ contour. {\it Right panel:} The same as the left panel, except that the slice of the $S^2$ function lies in the $M_{\ast}-\log{q(\rm H)}$ plane and the other two parameters are set to their optimal values of $T_{\rm eff}$=34,806 K and $\log{g}$=5.788.}
\label{chi2plots}
\end{figure*} 

Our approach to asteroseismology relies on a double optimisation procedure that first determines and quantifies the best match between the set of observed periodicities and those calculated for a given model, imposing certain restrictions on the mode identification if necessary (see Section 3.2). Subsequently, we identify  the model (or family of models) that can reproduce the observed periods most accurately within pre-defined limits of the 4-dimensional ($T_{\rm eff},\log{g},M_{\ast},\log{q(\rm H)}$) model parameter space. The latter is known as the ``optimal'' model, and corresponds to the absolute minimum of the goodness-of-fit merit function $S^2$ found by a dedicated optimisation code based on a hybrid Genetic Algorithm (GA, see \citealt{charp2008} for details). The merit function is given by 

\begin{equation}
S^2=\sum^n_{i=1}\frac{1}{\sigma}(P^i_{obs}-P^i_{th})^2
\end{equation}

where $P^i_{obs}$ is one of the $n$ periodicities observed, $P^i_{th}$ is the theoretical period that matches it best, and $\sigma$ is the global weight, defined as the inverse of the mode density of each model (i.e. the ratio of the width of the considered period window to the number of modes in that window). Including the global weight in the computation of $S^2$ removes, at least partially, possible large scale biases towards models with higher mode densities, which are usually found at the low surface gravity end of the parameter space explored. It is largely a technical choice, designed to prevent the GA code from allocating too many computational resources to a region of parameter space with unrealistic models that are in conflict with spectroscopic estimates of $\log{g}$.

\subsection{Search for the optimal model} 

We searched parameter space for an optimal model based on the 5 independent harmonic oscillations listed in Table \ref{freqs} as $f_1$ to $f_5$. For the reasons explained above, the rotationally split components for $f_3$ were not included, and the lower amplitude oscillations identified were deemed too insecure for the asteroseismological search. As has become standard in recent asteroseismological studies, the effective temperature was kept fixed at the spectroscopically determined value of $T_{\rm eff}$=34,806 K, while the other parameters were confined by limits designed to sandwich EC 09582-1137 in 3-dimensional $\log{g}-M_{\ast}-\log{q(\rm H)}$ parameter space. This yields the most accurate results because the $p$-mode frequencies in sdB stars are only weakly dependent on the effective temperature, and the latter can generally be determined more accurately from spectroscopy than asteroseismology. Moreover, it is well known that the period spectra computed for different models are subject to a degeneracy in $T_{\rm eff}-M_{\ast}$ space, implying that one of the two parameters must be constrained in order to be able to infer the other (see e.g. \citealt{charp2005a}). The boundaries set for the surface gravity cover a generous range centred on the spectroscopic estimate, with 5.70 $\leq \log{g} \leq$ 5.90. For the remaining two parameters we relied on stellar evolution theory and the possible ranges computed for various evolutionary scenarios \citep{han2002,han2003}, setting limits of 0.30 $\leq M_{\ast}/M_{\odot} \leq$ 0.7 and -5.20 $\leq \log{q(\rm H)} \leq$ -2.00. We computed modes from 50$-$350 s, which amply covers the observed period spectrum. 

\begin{table}[b]
\caption{Models that can account well for the five secure harmonic oscillations detected for EC 09582-1137.}
\label{models}
\centering
\begin{tabular}{@{\extracolsep{\fill}}c c | c c | c c}

\hline\hline

Rank & Period (s)& $\ell$ & $k$ & $\ell$ & $k$ \\

\hline 

$f_4$ & 169.10 & 2 & 2 & 2 & 1 \\
$f_5$ & 162.48 & 4 & 2 & 4 & 1 \\
$f_2$ & 151.24 & 0 & 2 & 1 & 2 \\
$f_3$ & 143.14 & 2 & 3 & 2 & 2 \\
$f_1$ & 136.00 & 0 & 3 & 0 & 2 \\
\hline
\multicolumn{2}{c}{} &\multicolumn{2}{c} {Model I} &\multicolumn{2}{c} {Model II} \\
\hline
\multicolumn{2}{c}{$S^2$} &\multicolumn{2}{c} {0.146} &\multicolumn{2}{c} {0.188} \\
\multicolumn{2}{c}{$\Delta P/P$} &\multicolumn{2}{c} {0.41 \%} &\multicolumn{2}{c} {0.57 \%} \\
\multicolumn{2}{c}{$\log{g}$} &\multicolumn{2}{c} {5.749} & \multicolumn{2}{c}{5.788} \\
\multicolumn{2}{c}{$M_{\ast}/M_{\odot}$} & \multicolumn{2}{c}{0.70} & \multicolumn{2}{c}{0.49} \\
\multicolumn{2}{c}{$\log{q(\rm H)}$} & \multicolumn{2}{c}{$-$3.34} &\multicolumn{2}{c} {$-$4.48} \\
\hline

\end{tabular}
\end{table}

\begin{table*}[t]
\centering
\caption{Period fit for the optimal model.}
\label{perfit}
\begin{tabular}{cccccccccccc}
\hline\hline

& & $P_{\rm obs}$ & $P_{\rm th}$ & $\sigma_I$& $\log E$ & $C_{kl} $ & $\Delta X/X$ & $\Delta P$  & Comments\tabularnewline
 $l$ & $k$ & (s) & (s) & (rad/s) & (erg) & & ($\%$) & (s) & \tabularnewline
\hline

0 & 4 & ...  & 103.0572 & $-$5.860 $\times$ $10^{-5}$ & 40.030 & ... & ... & ... & \tabularnewline
0 & 3 & ...  & 113.0118 & $-$2.057 $\times$ $10^{-5}$ & 40.575 & ... & ... & ... & \tabularnewline
0 & 2 & 136.0000 & 134.7030 &  $-$8.619 $\times$ $10^{-6}$  & 40.897 & ... & $+0.9537$ & $+1.2970$ & $\sharp$1 \tabularnewline
0 & 1 &  ... & 156.8309 &  $-$2.918 $\times$ $10^{-7}$  & 41.986 & ... & ... & ... & \tabularnewline
0 & 0 & ... & 172.1053 &  $-$1.004 $\times$ $10^{-7}$ & 41.956 & ... & ... & ... & \tabularnewline
 & & & & & & & & & \tabularnewline
1 & 5 & ... & 101.1261 &  $-$5.016 $\times$ $10^{-5}$  & 40.053 & 0.01330 & ... & ... & \tabularnewline
1 & 4 & ... & 110.5917 &  $-$3.274 $\times$ $10^{-5}$  & 40.357 & 0.01420 & ... & ... & \tabularnewline
1 & 3 & ... & 133.4155 &  $-$9.098 $\times$ $10^{-6}$ & 40.879  & 0.01446 & ... & ... & \tabularnewline
1 & 2 & 151.2400 & 151.1447 &  $-$5.241 $\times$ $10^{-7}$  & 41.840 & 0.03368 & $+0.0630$ & $+0.0953$  & $\sharp$2 \tabularnewline
1 & 1 & ... & 171.2927 &  $-$1.263 $\times$ $10^{-7}$  & 41.895 & 0.03368 & ... & ... & \tabularnewline
 & & & & & & & & & \tabularnewline
2 & 4 & ... & 108.0918 &  $-$5.047 $\times$ $10^{-5}$  & 40.152 & 0.01655 & ... & ... & \tabularnewline
2 & 3 & ... & 129.7607 &  $-$8.736 $\times$ $10^{-6}$  & 40.907 & 0.03740 & ... & ... & \tabularnewline
2 & 2 & 143.1400 & 142.1856 & $-$2.269 $\times$ $10^{-6}$  & 41.352 & 0.05828 & $+0.6667$ & $+0.9544$ & $\sharp$3 \tabularnewline
2 & 1 & 169.1000 & 169.9215 &  $-$1.664 $\times$ $10^{-7}$ & 41.827 & 0.02326 & $-0.4858$ & $-0.8215$ & $\sharp$4  \tabularnewline
2 & 0 & ... & 204.2292 &  $+$2.789 $\times$ $10^{-10}$  & 44.751 & 0.38900 & ... & ... & \tabularnewline
 & & & & & & & & & \tabularnewline
3 & 4 & ... & 105.9956 &  $-$6.152 $\times$ $10^{-5}$  & 40.045 & 0.01944 & ... & ... & \tabularnewline
3 & 3 & ... & 122.2176 &  $-$7.848 $\times$ $10^{-6}$  & 40.979 & 0.07823 & ... & ... & \tabularnewline
3 & 2 & ... & 136.9731 &  $-$6.207 $\times$ $10^{-6}$  & 40.999 & 0.03666 & ... & ... & \tabularnewline
3 & 1 & ... & 167.4583 &  $-$2.230 $\times$ $10^{-7}$  & 41.779 & 0.04495 & ... & ... & \tabularnewline
3 & 0 & ... & 181.5063 &  $+$1.734 $\times$ $10^{-9}$  & 43.114 & 0.19448 & ... & ... & \tabularnewline
3 & 1 & ... & 317.2987 &  $+$2.217 $\times$ $10^{-12}$  & 46.981 & $-$0.01281 & ... & ... & \tabularnewline
 & & & & & & & & & \tabularnewline
4 & 4 & ... & 104.0020 &  $-$6.328 $\times$ $10^{-5}$  & 40.002 & 0.02428   & ... & ... & \tabularnewline
4 & 3 & ... & 115.8280 &  $-$1.270 $\times$ $10^{-5}$  & 40.777 & 0.06850   & ... & ... & \tabularnewline
4 & 2 & ... & 134.7884 &  $-$8.039 $\times$ $10^{-6}$  & 40.907 & 0.02637   & ... & ... & \tabularnewline
4 & 1 & 162.4800 & 163.5995 &  $-$2.774 $\times$ $10^{-7}$  & 41.790 & 0.07945 & $-0.6890$ & $-1.1195$ & $\sharp$5 \tabularnewline
4 & 0 & ... & 175.1735 &  $-$1.549 $\times$ $10^{-8}$  & 42.383 & 0.100063 & ... & ... & \tabularnewline
4 & 1 & ... & 271.3005 &  $+$6.164 $\times$ $10^{-12}$  & 46.522 & $-$0.03263 & ... & ... & \tabularnewline
\hline

\end{tabular}
\end{table*}

It is important to mention that, although we do not have access to a priori mode identification from e.g. multi-colour photometry or time-series spectroscopy, we place some constraints on the degree index of the observed modes in what follows. This is based entirely on mode visibility arguments, and assumes a similar intrinsic amplitude for all the modes considered. As illustrated  in Fig. 9 of \citet{randall2007}, mode visibility when integrated over the visible disk of the star generally decreases as the degree index increases for the vast majority of inclinations of the pulsational axis compared to the line of sight, radial and dipole modes dominating the amplitude hierarchy for an inclination angle $i\lesssim$ 70$^\circ$. It has moreover been shown that because of their specific surface geometry, modes with $\ell$=3 have an extremely low visibility in the optical domain due to cancellation effects \citep{randall2005}. Therefore, we consider only modes with $\ell$ = 0,1,2,4 in our asteroseismological analysis. In an initial search, the identification of a given observed periodicity was left open within these constraints, however this produced a large number of models that were able to account for the frequencies observed quite well. Given that the two highest amplitude oscillations $f_1$ and $f_2$ completely dominate the observed period spectrum, we thought it justified to limit them to radial or dipole modes, while leaving the mode identification of the lower amplitude modes open to $\ell$ = 0,1,2,4. 

In the search domain specified, this approach identified two families of models able to reproduce the observed periodocities within the constraints in terms of $p$-modes\footnote{A third model was identified, however it contained mixed modes, i.e. modes that show characteristics of both $p-$ and $g-$modes. Since these propagate into the deeper layers of the star, they cannot be accurately modelled with the 2G models. Therefore, this model could be excluded.}. These are listed in Table \ref{models}, together with the mode identification inferred. It is apparent that both models are consistent with the spectroscopic estimate of $\log{g}$ within the errors ($T_{\rm eff}$ was fixed to the spectroscopic value of 34,806 K), although Model II lies closer to the value measured. The two models feature a plausible mode identification, the lower amplitude modes being associated with modes of relatively high degree index. In terms of the quality of the period match Model I fares slightly better than Model II, but the fit is acceptable in both cases. 

\begin{figure}[t]
\centering
\includegraphics[width=7.0cm,bb=100 150 500 680]{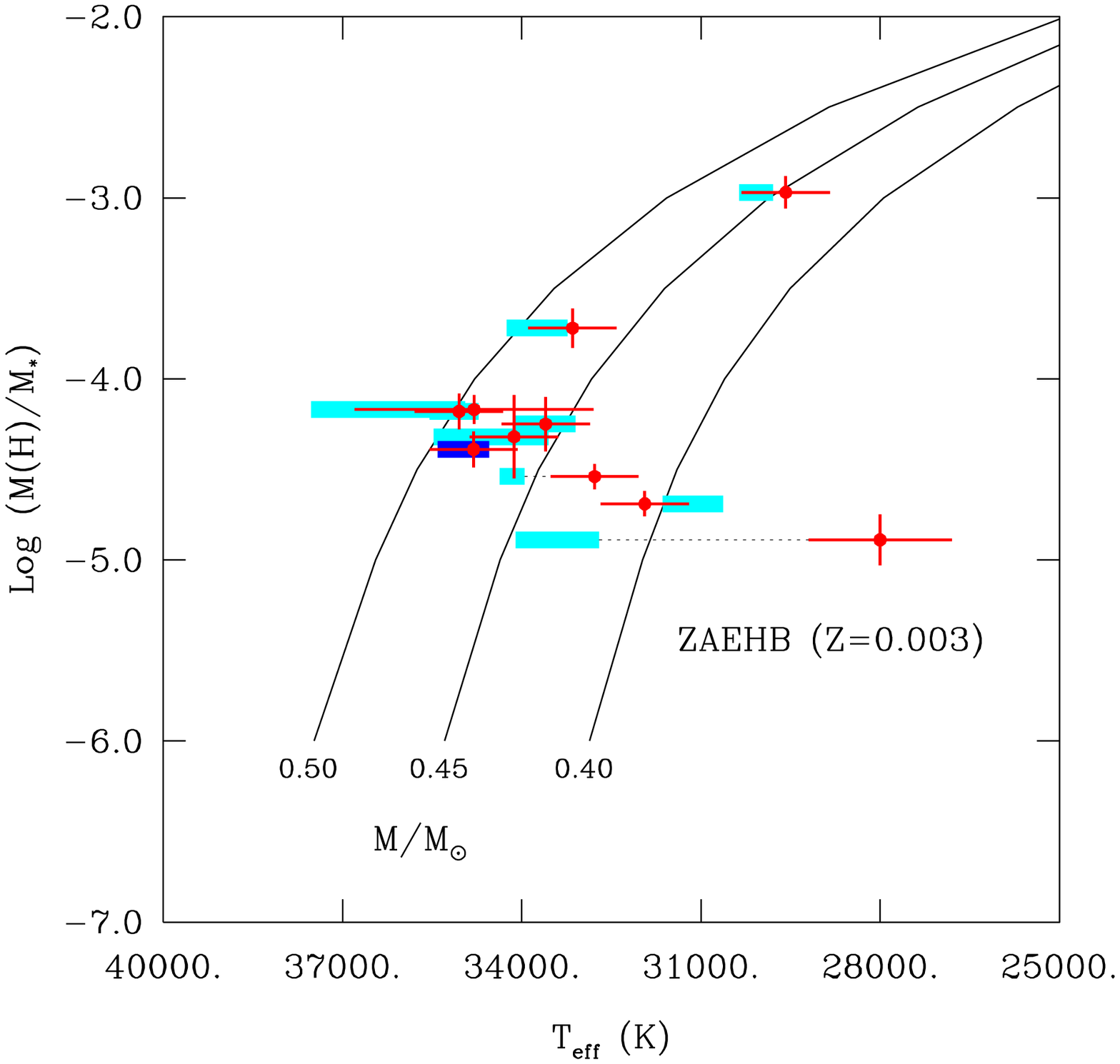}
\caption{Expected (solid curves) and observed (dots with error bars) correlations between the effective temperature, the fractional hydrogen envelope mass, and the total mass (thick bars projected onto the theoretical mass relations) for the 10 sdB pulsators so far submitted to asteroseismology. EC 09582-1137 is indicated by the darker bar. }
\label{evol}
\end{figure}

Looking at the model parameters more closely, we find that Model I has a rather high mass, at the edge of the search limits defined. This alone does not exclude it as a viable model, since masses significantly higher and lower than the canonical value are possible according to binary evolution calculations \citep{han2002,han2003,yu2009}. However, the {\it relative} values of $M_{\ast}$ and $\log{q(\rm H)}$ derived are implausible from an evolutionary point of view. Regardless of how it is formed, an sdB star is expected to start its helium burning lifetime on a zero-age EHB (ZAEHB) specified by its total mass and initial metallicity. Its exact location on that ZAEHB is determined by the thickness of its hydrogen rich envelope. So at $t$=0, for a given metallicity, a precise correlation exists between the total mass, the thickness of the H shell, and the effective temperature \citep[for details see][]{fontaine2006}. This correlation is shown in Fig. \ref{evol} for total stellar masses of $M_{\ast}/M_{\odot}$=0.40, 0.45 and 0.50 assuming a metallicity of $Z$=0.003. While the high mass of $M/M_{\ast}$=0.70 predicted for Model I is not shown, it is clear from an extrapolation of the plot that, given the spectroscopic temperature $T_{\rm eff}\sim$ 35,000 K one would expect a much more massive envelope than that indicated in Table \ref{models}, with $\log{M(\rm H)/M_{\ast}}\gtrsim -$2. 

The relations displayed in Fig. \ref{evol} are assumed to be valid to first order during most of the He-burning lifetime of sdB stars, but break down for individual objects that have already evolved away from the ZAEHB (see the unusual case of Balloon 090100001 presented by \citealt{val2008b}). The locations on the plot of the stellar parameters derived for the targets so far submitted to asteroseismology are in accordance with the theoretical relations, nicely illustrating the strong internal consistency that exists between the derived parameters. The one exception (marked by the outlying cross) is for Balloon 090100001, a supposedly more evolved object. Given that EC 09582-1137 shows both atmospheric and pulsational characteristics typical of EC 14026 pulsators, it makes sense to assume that the ZAEHB relations hold true, allowing us to exclude Model I as physically implausible.

We thus retain Model II as the optimal model, and show the behaviour of the merit function in its vicinity in parameter space in Fig. \ref{chi2plots}. It is evident that the atmospheric parameters inferred agree perfectly with the spectroscopic estimates for $\log{g}$ and $T_{\rm eff}$. As predicted, the solution is far more sensitive to the surface gravity than the effective temperature because of the stronger period dependence. In $M_{\ast}-\log{q(\rm H)}$ space, the solution is well defined for both parameters, as is to be expected from their strong signature on the frequency spectrum. Examining the mode identification, we see that the observed amplitude of a periodicity decreases with increasing degree index $\ell$, which makes the solution plausible from a mode visibility point of view. Of course, the mode identification was partly fixed from the outset, as $f_1$ and $f_2$ were required to be radial or dipole modes. However, it was not obvious that the lower amplitude modes would be associated with the less visible $\ell$=2 and $\ell$=4 modes. This is particularly interesting for the case of $f_3$, which from the available data shows a triplet structure attributed to rotational splitting. If it is indeed a quadrupole mode as suggested by our mode identification, we should see a quintuplet given a long enough time baseline. With the current data set, we are presumably able to detect only the outer $m$ = $\pm$2 components, and cannot resolve the $m$ = $\pm$1 peaks. This could explain why we do not observe rotational splitting for the higher amplitude $f_2$ periodicity, while the multiplet components of the weak $f_4$ and $f_5$ peaks are probably lost in the noise. A longer data set would shed light on this, and could be exploited to confirm the mode identification and thus the validity of our optimal model.

\begin{figure}[t]
\centering
\includegraphics[width=7.0cm,bb=150 200 500 650]{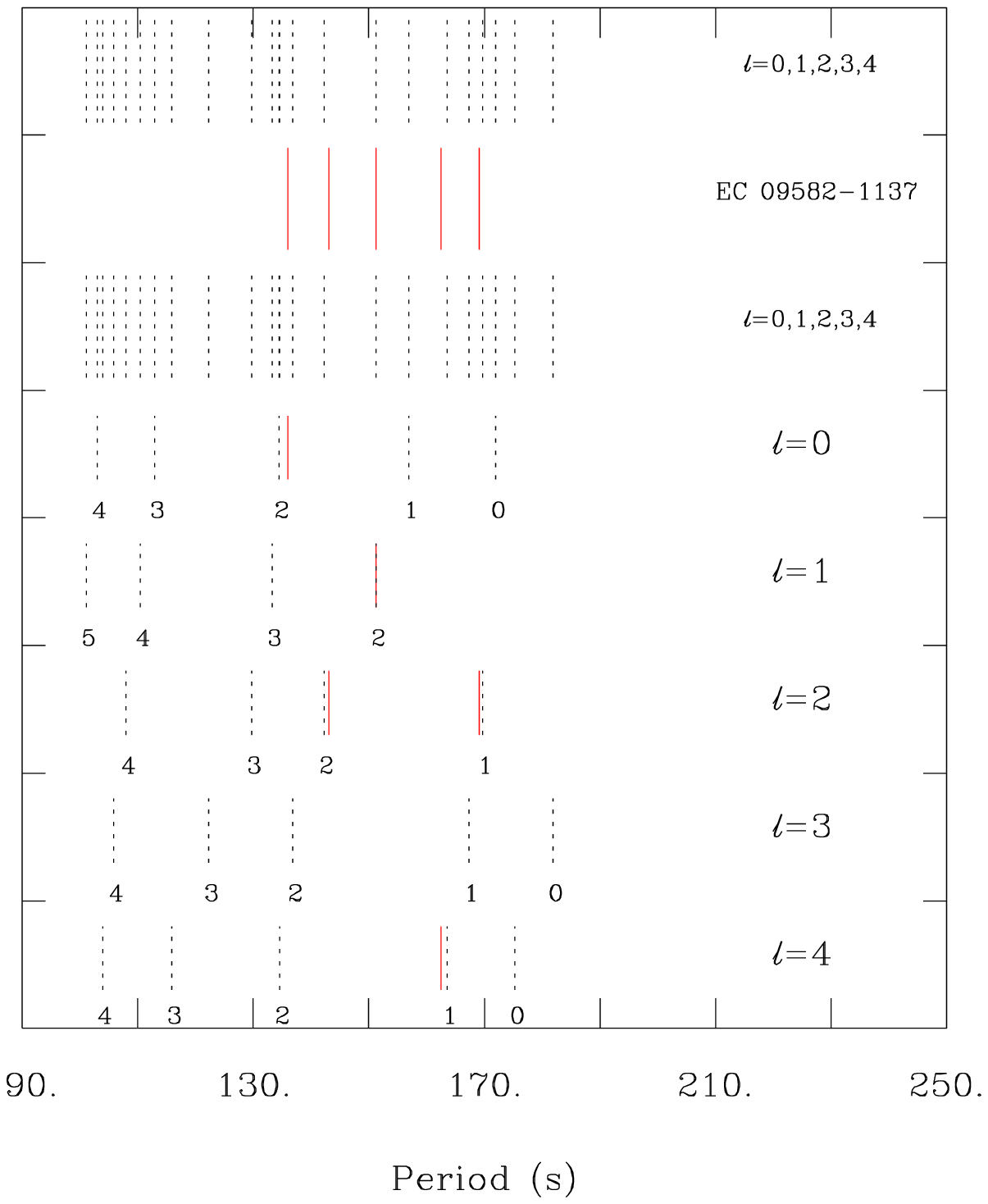}
\caption{Comparison of the observed period spectrum of EC 09582-1137 (continuous line segments) with the theoretical period spectrum of the optimal model (dotted line segments) in the 100$-$200 s range for degree indices $\ell$=0,1,2,3,4. Note that al the theoretical modes plotted are predicted to be excited and correspond to low-order $p$-modes. The radial order $k$ is indicated below each segment. 
}
\label{compper}
\end{figure} 

\subsection{Period fit and structural parameters for EC 09582-1137} 

The optimal model can account for the 5 observed periods to within 0.57 \% (corresponding to an absolute period dispersion of $\Delta$P=0.833 s), which is quite typical for the asteroseismological analyses carried out to date for EC 14026 stars. Details on the optimal model period spectrum can be found in Table \ref{perfit}, where we list the periods computed together with their degree index $\ell$ and radial order $k$, kinetic energy $\log{E}$, rotation coefficient $C_{kl}$ and stability coefficient $\sigma_I$. When the latter is negative, the oscillation is predicted to be excited from non-adiabatic theory. The fact that this is the case for all modes assigned to an observed periodicity constitues an important a posteriori consistency check for the validity of our optimal model. 

The periods observed for EC 09582-1137 are shown next to their theoretical counterparts, and the absolute and relative dispersion is indicated for each. A posteriori it is interesting to note that the two lower amplitude periodicities detected at 132.83 s and 135.72 s (see Section 2.2) can be respectively matched to the $\ell$=1, $k$=3 and the $\ell$=4, $k$=2 mode without a significant deterioration of the merit function occurring. In fact, if we repeat our asteroseismic search including these extra two frequencies, we find an absolute minimum in the merit function at values that are compatible with the adopted optimal model within the errors. The absolute period dispersion then increases slightly, but the relative period fit remains unchanged. 

In order to better illustrate the quality of the period fit, we include a graphical representation in Fig. \ref{compper}. As has been noted in previous studies, the dispersion of the period fit is an order of magnitude larger than the measurement uncertainty on the periods. This is attributed to inadequacies in the 2G models, which can undoubtedly be improved upon. Work on the so-called {\it third-generation} models \citep{brassard2008} is on-going, and should help address at least some of the outstanding issues. 

The solution obtained from asteroseismology leads to a natural determination of the three variable input parameters $\log{g}$, $M_{\ast}$ and $\log{q(\rm H)}$, while the effective temperature is known from spectroscopy. On the basis of these quantities we can derive a number of secondary parameters: the stellar radius $R$ (as a function of $g$ and $M_{\ast}$), the luminosity $L$ (as a function of $T_{\rm eff}$ and $R$), the absolute magnitude $M_V$ (as a function of $g$, $T_{\rm eff}$ and $M_{\ast}$ in conjunction with detailed model atmospheres) and the distance from Earth $d$ (as a function of apparent magnitude and $M_V$; we assumed $V$=15.26$\pm$0.05). As detailed in Section 2, the rotation period can tentatively be set to $P_{rot}\gtrsim$ 2.63 d from the rotational splitting observed.
 The derived parameters are listed in Table \ref{modelfit} together with the internal 1-$\sigma$ errors computed according to the recipe given in \citet{charp2005b} for the three input parameters, and propagated through for the secondary quantities. All the errors are purely statistical, and almost certainly underestimate the true uncertainties arising from systematic effects.

\begin{table}
\caption{Inferred structural parameters for EC 09582-1137}
\label{modelfit}
\begin{tabular}{lrclc}
\hline\hline

Quantity & \multicolumn{3}{c}{Estimated Value} & \tabularnewline

\hline\hline

 $T_{\rm eff}$ (K)$^{\dag}$ & $34806$ & $\pm$ & $230$ & ($0.66 \%$) \tabularnewline
 $\log g$ & $5.788$ & $\pm$ & $0.004$ & ($0.07 \%$) \tabularnewline
 $M_*/M_{\odot}$ & $0.485$ & $\pm$ & $0.011$ & ($2.27 \%$) \tabularnewline
 $\log (M_{\rm env}/M_*)$ & $-4.39$ & $\pm$ & $0.10$ & ($2.23 \%$) \tabularnewline
 & \tabularnewline
 $R/R_{\odot}$ ($M_*$, $g$) & $0.147$ & $\pm$ & $0.002$ & ($1.57 \%$) \tabularnewline
 $L/L_{\odot}$ ($T_{\rm eff}$, $R$) & $28.6$ & $\pm$ & $1.7$ & ($5.79 \%$)
\tabularnewline
 $M_V$ ($g$, $T_{\rm eff}$, $M_*$) & $4.44$ & $\pm$ & $0.05$ & ($1.13 \%$)
\tabularnewline
 $d$ ($V$, $M_V$) (pc) & $1461$ & $\pm$ & $66$ & ($4.52 \%$) \tabularnewline
 %$P_{\rm rot}$ (day) & $0.0$ & $\pm$ & $0.0$ & ($nan \%$) \tabularnewline
 %$V_{\rm eq}$ ($P_{\rm rot}$, $R$) (km/s) & $inf$ & $\pm$ & $nan$ & ($nan \%$)
%\tabularnewline
% $i$ ($\deg$) & $0.0$ & $\pm$ & $0.0$ & ($nan \%$) \tabularnewline

\hline

{\footnotesize $^{\dag}$ From spectroscopy}
\end{tabular}
\end{table}

\section{Conclusion}

In this paper we presented the asteroseismological analysis for a tenth rapidly pulsating subdwarf B star, the relatively recently discovered EC 09582-1137. On the basis of $\sim$ 30 hours of SUSI2 time-series photometry we uncovered 5 independent harmonic oscillations as well as two periodicities interpreted as the rotationally split components of a frequency multiplet. The first harmonic of the dominant oscillation was additionally uncovered, albeit at an amplitude below the imposed detection threshold. We also obtained a high S/N low resolution spectrum in order to infer the target's atmospheric parameters with some accuracy. Using the observed oscillations as input, we conducted an astero{-}seismic search in parameter space to find the model that could optimally account for the period spectrum of EC 09582-1137 . Given that we had only five observed periodicities as input we placed some constraints on the mode identification, requiring the two dominant pulsations to correspond to radial or dipole modes. It was then possible to isolate a well-defined optimal model that could reproduce all the observed oscillations simultaneously to within 0.6\%. As has become standard in the asteroseismology of EC 14026 stars, the solution is in accordance with the spectroscopic estimates of the atmospheric parameters, and all the observed modes are predicted to be excited from non-adiabatic calculations.

Judging by the optimal model identified, EC 09582-1137 appears to be a very typical member of the EC 14026 pulsator class in terms of fundamental parameters. The values of $\log{g}$ and $T_{\rm eff}$ place it in the middle of the instability strip, and its mass is close to the canonically expected value. Our target shows no spectroscopic indication of a companion, and appears to be rotating slowly, with a rotation period of the order of a few days. Given that sdB stars in close binary systems with $P_{bin}\lesssim$ 0.6 days have been shown to exhibit binary-synchronous rotation \citep{geier2008,val2008a}, such objects should display rotational splitting corresponding to the binary period in their frequency spectra. Despite a sufficient time baseline, we find no indication of such a short rotation period in our photometry, and conclude that, if EC 09582-1137 forms part of a binary system, the latter has a period longer than $\sim$ 0.6 days. It is therefore a good candidate for a single star or a component of a relatively wide binary with an unseen companion, but this remains to be clarified from radial velocity measurements.

The fact that the work presented here constitutes the tenth asteroseismological analysis of a rapidly pulsating subdwarf B star has in our opinion demonstrated beyond doubt the basic validity of our 2G models and the GA approach. Of course, there is still room for improvement on the theoretical side. Shortcomings in the current models are indicated by the relatively poor dispersion of the asteroseismic period fit when compared to the observational accuracy of the periodicities. The deficiencies of the 2G sdB star models become even more apparent when studying the slowly pulsating hot subdwarfs. These oscillators exhibit gravity modes that probe deeper inside the star than pressure modes, and are as such sensitive to the inner layers not modelled accurately by the 2G models. Work on so-called third generation models incorporating nuclear processes is ongoing, and is hoped to improve the accuracy of asteroseismic solutions.

On the observational side, much remains to be done. Although we were allocated 5 nights of observations on a medium-size telescope using a highly sensitive imager, we detected only 5 independent harmonic oscillations. As a result of this, we were forced to make some assumptions as to the mode identification based purely on visibility arguments. A longer time baseline allowing us to analyse the rotational signature and/or multi-colour information would yield partial mode identification, making such assumptions unnecessary. Moreover, an increased sensitivity would provide additional secure frequencies, strengthening the robustness of the asteroseismic solution. Possible avenues for future observational campaigns include space telescopes designed for asteroseismology such as Kepler or Corot, and networks of small telscopes providing world-wide coverage. An alternative is to obtain shorter high quality data sets on large telescopes, however there is currently little instrumentation available for fast time-series imaging. We remain hopeful that this will change in the near future, e.g with the second-generation suite of instruments on the VLT or the new instrumentation for GTC.

\begin{acknowledgements}
S.K.R. would like to thank the ESO La Silla staff, in particular SUSI2 instrument scientist Alessandro Ederoclite for their support, motivation and friendly welcome. V. V. G. acknowledges grant support from the Centre National d'Etudes spatiales. G.F. acknowledges the contribution of the Canada Research Chair Programme. We would also like to thank an anonymous referee for useful comments and suggestions. Sadly, the observations reported here were among the last ever obtained with SUSI2 as the instrument was decommissioned only a few days after the end of our run. EMMI has since followed. May they rest in peace.
\end{acknowledgements}

\bibliographystyle{aa}
\bibliography{12576}

\end{document}